\documentclass{sf2a-conf2013}
\usepackage{graphicx}
\usepackage{hyperref}
\usepackage[]{natbib}  
\usepackage{epstopdf}

\def\BibTeX{{\rm B\kern-.05em{\sc i\kern-.025em b}\kern-.08em
    T\kern-.1667em\lower.7ex\hbox{E}\kern-.125emX}}
\bibpunct{(}{)}{;}{a}{}{,}  

\begin{document}

\TitreGlobal{SF2A 2013}


\title{Characterizing small planets transiting small stars with SPIRou}

\runningtitle{SPIRou characterizing small planets transiting small stars}

\author{A.~Santerne}\address{Centro de Astrof\'isica, Universidade do Porto, Rua das Estrelas, 4150-762 Porto, Portugal}

\author{J.-F.~Donati}\address{UPS-Toulouse / CNRS-INSU, Institut de Recherche en Astrophysique et Plan\'etologie (IRAP) UMR 5277, Toulouse, FÐ31400 France}

\author{R.~Doyon}\address{D\'epartement de physique and Observatoire du Mont M\'egantic, Universit\'e de Montr\'eal, C.P. 6128, Succursale Centre-Ville, Montr\'eal, QC H3C 3J7, Canada}

\author{X.~Delfosse}\address{UJF-Grenoble 1/CNRS-INSU, Institut de Plan\'etologie et d'Astrophysique de Grenoble (IPAG) UMR 5274, 38041 Grenoble, France}

\author{E.~Artigau$^{3}$}

\author{I.~Boisse$^{1}$}

\author{X.~Bonfils$^{4}$}

\author{F.~Bouchy}\address{Aix Marseille Universit\'e, CNRS, LAM (Laboratoire d'Astrophysique de Marseille) UMR 7326, 13388 Marseille, France}

\author{G.~H\'ebrard$^{6,}$}\address{Institut dÕAstrophysique de Paris, CNRS, Universit\'e Pierre et Marie Curie, 98bis Bd. Arago, 75014 Paris, France}\address{Observatoire de Haute-Provence, CNRS/OAMP, 04870 Saint-Michel-l'Observatoire, France}

\author{C.~Moutou$^{5,}$}\address{CNRS, Canada-France-Hawaii Telescope Corporation, 65-1238 Mamalahoa Hwy., Kamuela, HI 96743, USA}

\author{S.~Udry}\address{Observatoire de Gen\`eve, Universit\'e de Gen\`eve, 51 chemin des Maillettes, Sauverny 1290, Switzerland}

\author{the SPIRou science team}




\setcounter{page}{237}


\maketitle


\begin{abstract}
SPIRou, a near infrared spectropolarimeter, is a project of new instrument to be mounted at the Canada France Hawaii Telescope in 2017. One of the main objectives of SPIRou is to reach a radial velocity accuracy better than 1 m.s$^{-1}$ in the YJHK bands. SPIRou will make a cornerstone into the characterization of Earth-like planets, where the exoplanet statistics is very low. This is even more true for planets transiting M dwarfs, since only 3 low-mass planets have been secured so far to transit such stars. We present here all the synergies that SPIRou will provide to and benefit from photometric transit-search programs from the ground or from space (\textit{Kepler}, \textit{CHEOPS}, \textit{TESS}, \textit{PLATO 2.0}). We also discuss the impact of SPIRou for the characterization of planets orbiting actives stars.
\end{abstract}



\section{Introduction}
As they pass in front of their host star, transiting exoplanets are providing us numerous key information to understand the diversity of planets in the galaxy. For these planets, it is possible to determine their mass and radius and thus their bulk density, needed to determine their nature (rocky, gazeus, water-world, etc\dots) and model their internal structure as well as their orbital configuration (orbital period, eccentricity, obliquity). This provides strong constraints to planet's formation, migration and evolution theories. Last but not least, transiting exoplanets are today the only targets to explore atmospheric composition from transmission spectroscopy during the transit.\\ 

\begin{figure}[ht!]
 \centering
 \includegraphics[width=0.7\textwidth,clip]{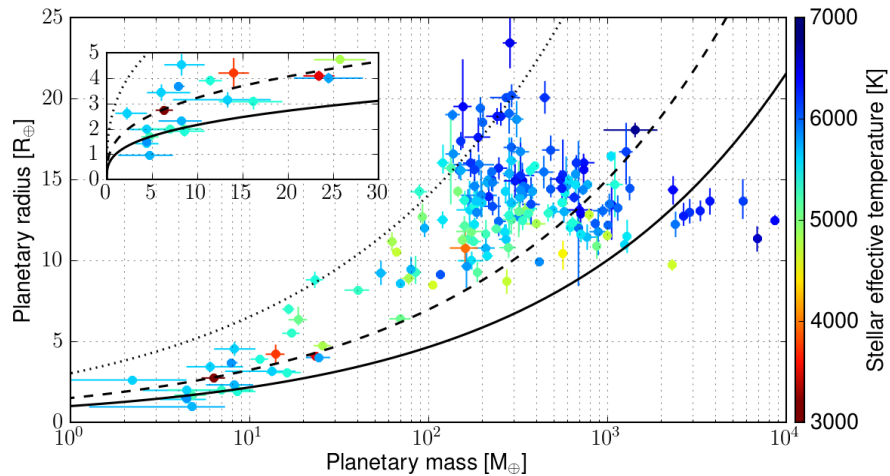}
  \caption{Mass-radius diagram of transiting exoplanets discovered so far. The colors display the effective temperature of the host stars. The inset is a zoom to the transiting super-Earths. Only a few small planets have been characterized around small stars (here displayed in red). The solid, dashed and dotted lines display the density of the Earth, Neptune and of 0.2 g.cm$^{-3}$ (respectively).}
  \label{santerne2:fig1}
\end{figure}

Nearly 300 transiting planets have been confirmed so far with only a handful of terrestrial or very-low mass planets having an accurate determination of mass and radius (see Figure \ref{santerne2:fig1}). To explore this regime of low-mass planets, small stars like M dwarfs are the most favorable for detection since they allow larger signals in both photometry (the depth of the transit scales as 1/R$_{\star}^{2}$) and radial velocity (hereafter RV; the amplitude of the RV variation scales as 1/M$_{\star}^{2/3}$). For example, for a planet in the habitable zone, the RV signal is seven times larger around a M dwarf than around a solar-type star (two effects combined: the habitable zone is closer to the stars and the mass of the star is lighter). \\

Since atmospheric characterization primarily requires deep transits on the one hand, and bright stars on the other hand (in the nIR, where absorption from atmospheric molecules mostly concentrates), M dwarfs are therefore optimal targets for this quest. Detailed simulations show that atmospheric components of Earth-like extrasolar planets will produce a detectable signal for planets around M dwarfs planets using JWST/NASA and/or ELTs, but not for similar planets around Sun-like stars \citep[e.g.][]{2011A&A...529A...8R,2012ExA....34..311T}. Today, only a handful of very-bright transiting systems have been discovered. A prime goal of astronomy is to discover other Earths and super-Earths whose atmosphere can be scrutinized and characterized with space missions (such as JWST) in the next decade, including the search for biomarkers.\\

Dedicated searches for planets around M dwarfs in radial velocity have been performed for years \citep[e.g.][with the HiReS and HARPS spectrographs]{1998A&A...338L..67D,1998ApJ...505L.147M, 2001ApJ...556..296M} unveiling $\sim$ 10 very low-mass planets \citep[e.g.][]{2005ApJ...634..625R}, including super-earths in the habitable zone of Gl~581Ê\citep{2009A&A...507..487M}, Gl~667C \citep{2013A&A...553A...8D, 2013arXiv1307.6984F} and Gl~163 \citep{2013A&A...556A.110B}. Very importantly, these surveys also revealed that small planets are very common around small stars, and that, $41\%^{_{+54\%}}_{^{-13\%}}$ of the M dwarfs harbor a super-earth planet in the habitable zone \citep{2013A&A...549A.109B}. Among the planets found around M dwarfs in RV surveys, only two have been found to transit their host star: GJ~436~b \citep{2004ApJ...617..580B, 2007A&A...472L..13G} and GJ~3470~b \citep{2012A&A...546A..27B}.  The Mearth ground-based photometric survey dedicated to M dwarfs reported a third super-Earth with measured mass and radius: GJ~1214~b \citep{2009Natur.462..891C}. All of them present a bulk density close to Neptune's density (see Fig. \ref{santerne2:fig1}). A giant planet transiting a M dwarf has been characterized among the \textit{Kepler} candidates by \citet{2012AJ....143..111J}, totalizing only four fully-characterized planets known to transit M dwarfs.\\

To improve the statistics on this population of planets and to discover the prime targets for mission dedicated to extra-solar planet's atmospheric characterization, it is very important to characterize many more planets transiting nearby M dwarfs. Since M dwarfs are quite faint in the optical but luminous in the infrared, such studies are much more efficient with infrared facilities.

\section{The SPIRou spectrograph}
The SPIRou spectrograph (SpectroPolarim\`etre Infra-Rouge) is a project for a near infrared spectropolarimeter that will be mounted at the 3.6-m Canada France Hawaii Telescope (CFHT) in 2017. The main goals of this unique spectropolarimeter will be both to search for habitable exo-Earths orbiting low-mass \& very-low mass stars using high-accuracy RV (better than 1 m.s$^{-1}$) and to explore the impact of magnetic fields on star \& planet formation, by detecting magnetic fields of various types of young stellar objects and by characterizing their large-scale topologies \citep{2011ASPC..448..771A}. This spectropolarimeter will be a high-resolution (R $\sim$ 75 000) stabilized spectrograph covering a spectral range between 0.98 $\mu$m and 2.35 $\mu$m (i.e. YJHK photometric bands) with a large throughput (up to 15\%). Thanks to its large throughput, SPIRou will be able to provide spectrum with a signal-to-noise ratio (SNR) of $\sim$ 110 per pixel in one hour on a star of magnitude J = 12 or K = 11. SPIRou is expected to reach a photon noise of 1m.s$^{-1}$ with a SNR of 160. Therefore SPIRou will be the best instrument for RV studies of M dwarfs, especially in the context of transiting planet surveys.

\section{Current and future transiting planet surveys}

Current and future (optical or infrared) photometric surveys are targeting M dwarfs to find new transiting planets. Ground-based nIR spectroscopy is essential in this context: spectroscopy is indeed mandatory to establish the planetary nature of all transiting objects detected around low-mass dwarfs through photometric monitoring \citep[and discard false detections, e.g., caused by background eclipsing binaries ;][]{2012A&A...545A..76S} and to measure their mass from the RV amplitude of their host dwarfsÕ orbital motion. SPIRou will also detect planets in a dedicated M-dwarf RV survey, that will be after search for transits by ground- or space-based observatories.

\subsection{\textit{Kepler}}

\begin{figure}[ht!]
 \centering
 \includegraphics[width=0.7\textwidth,clip]{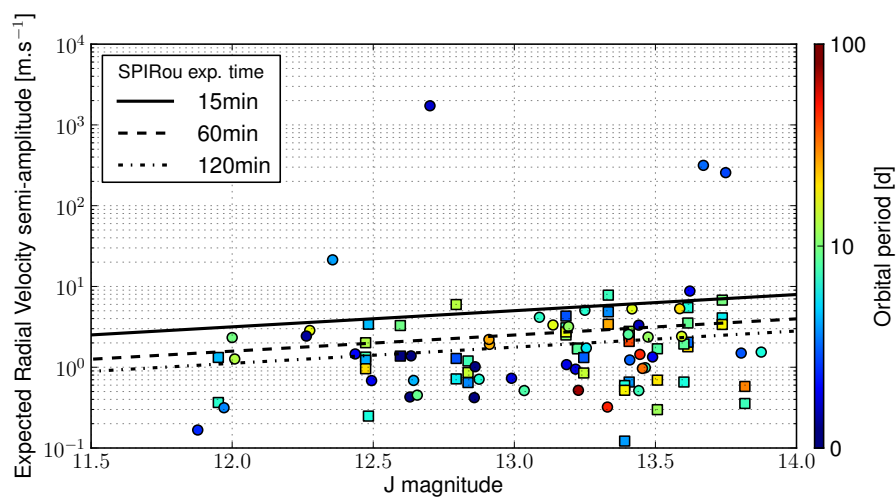}
  \caption{Expected radial velocity semi-amplitude of the 95 KOIs from \citet{2013ApJ...767...95D} as function of the J magnitude of their host star. The color of the mark indicates the orbital period of the candidate, while their shape indicates the multiplicity of the system: circles for single-planet candidates and squares for multiple-planetary systems. The black solid, dashed and dot-dashed lines indicate the RV accuracy that SPIRou will reach in 15 minutes, 1 hour and 2 hours (respectively) of exposure time.}
  \label{santerne2:fig2}
\end{figure}

The \textit{Kepler} space telescope \citep{2009Sci...325..709B} detected 95 planet candidates (the so-called Kepler Objects of Interest, i.e. KOIs) transiting M dwarfs \citep{2013ApJ...767...95D} with more than 2000 planet candidates \citep{2013ApJS..204...24B} transiting FGK stars. These KOIs present radii in the range 0.5 R$_{\oplus}$ -- 17 R$_{\oplus}$ and orbital periods in the range 0.5 -- 82 days. Most of them have a radius similar to the one of the Earth. Assuming bulk densities from the solar system planets (Neptune's density for planet candidates larger than 2.5 R$_{\oplus}$ and the Earth density for those smaller than 2.5 R$_{\oplus}$), it is possible to estimate their RV semi-amplitude. Figure \ref{santerne2:fig2} displays this expected RV amplitude, as function of the J magnitude of the host star. The majority of these KOIs are expected to present a RV semi-amplitude larger than 1 m.s$^{-1}$. Radial velocity follow-up have been initiated with optical facilities (e.g. with SOPHIE, HiReS, HARPS-N) but no low-mass planet has been characterized yet. Observations with an infrared spectrograph, like SPIRou, would be much more efficient. If SPIRou was already built, it would already be able to characterize new small planets transiting the smallest \textit{Kepler} stars. However, since the \textit{Kepler} targets are faint, this would required a lot of telescope time.

\subsection{Ground-based dedicated M dwarf surveys}

Several ground-based photometric surveys are only targeting M dwarfs to detect small transiting planets. For example, the Mearth observatory was used to discover the transiting mini-Neptune GJ~1214~b \citep{2009Natur.462..891C}. Other similar facilities are in preparation, such as ExTrA, Apache and Speculoos. For example, the ExTrA (Exoplanets in Transit and their Atmospheres) infrared observatory that will start observations in 2015 and is expected to discover nearly 50 new small planets (down to 0.5 R$_{\oplus}$) transiting bright M dwarfs (Bonfils, private communication). The infrared spectrograph SPIRou will be able to characterize these new transiting planets with a better efficiency than other optical facilities. Moreover, those ground-based observatory will be able to perform a photometric follow-up of the new planets that SPIRou will detect as part of the RV survey.

\subsection{\textit{TESS}}

\textit{TESS} (Transiting Exoplanet Survey Satellite) is a all-sky space-based photometric survey of all stars brighter than V magnitude of 12 \citep{2010AAS...21545006R} and M-dwarfs until V=13 (Charbonneau, private communication) that will be launched in 2017. This survey will therefore include numerous M dwarfs. \textit{TESS} is expected to detect more than 300 earth and super-earth transiting bright stars\footnote{\url{http://science.nasa.gov/media/medialibrary/2013/04/22/secure-RICKER-TESS_NASA_APS_17Apr2013_NoVideo_v4-1.pdf}}. The vast majority of them will orbit around M dwarfs due to the deeper signal for smaller stars. Since (i) most Earths and super-Earths detected with TESS will orbit around M dwarfs, and (ii) less than $\sim$ 30\% of them will be accessible to optical RV follow-up \citep{2009PASP..121..952D}, SPIRou will be the best RV instrument to monitor in the near infrared the $\sim$ 150 best candidates visible from CFHT (assuming $\sim$ 60 visits per star and spectra SNR of $\sim$ 160 per visit, this observational effort will represent a total of 150 CFHT nights).

\subsection{\textit{PLATO 2.0}}

\textit{PLATO 2.0} (PLAnetary Transit and Oscillations of stars; Rauer et al., 2013) is a M3 mission candidate to the ESA Cosmic Vision program. If selected, \textit{PLATO 2.0} will operate from 2024 to 2031+. Its main objective is to detect transiting planets out to the habitable zone of bright Sun-like stars and to characterize the host star simultaneously through asteroseismology. Since they orbit bright stars, those planets will be more easily characterizable with ground-based spectrographs than the much fainter \textit{Kepler} targets. With \textit{PLATO~2.0}, it will be possible de measure the density of planets with an unprecedented accuracy. Like \textit{TESS}, \textit{PLATO 2.0} will observe thousands of bright M dwarfs but will deeper probe their planet population, towards smaller planets and longer orbital periods (even longer than the habitable zone of early-M dwarfs). To characterize these unique planets that only \textit{PLATO 2.0} will be able to find, infrared spectrographs like SPIRou are needed.

\subsection{Synergies with \textit{CHEOPS}}

\textit{CHEOPS} \citep[CHaracterising ExOPlanets Satellite][]{2013EPJWC..4703005B} is the first S-class mission selected by ESA in its Cosmic Vision Program. The objective of this space mission is to detect new transiting planets around bright stars by performing a photometric follow-up of known planets detected by RV surveys. In operation between 2017 and 2021, \textit{CHEOPS} will be able to observe photometrically the first planets detected by SPIRou as part of its RV survey.

\section{SPIRou and stellar activity}
Stellar activity is one of the main limitation to the characterization of transiting planets \citep[e.g.][]{2011ApJ...742...59H}. The case of CoRoT-7 \citep{2009A&A...506..287L} is a good illustration of this limitation in the context of the characterization of small planets around active stars \citep[for a complete view of the saga about the mass of CoRoT-7~b, see:][]{2009A&A...506..303Q, 2010A&A...520A..53L, 2010A&A...520A..93H, 2011MNRAS.411.1953P, 2011A&A...528A...4B, 2011A&A...531A.161F, 2011ApJ...743...75H}. To better disentangle stellar activity from the planetary signal in CoRoT-7~b, new data from \textit{CoRoT} and HARPS have been obtained simultaneously and will be discussed in Haywood et al. (submitted), Barros et al. (in prep.), Hatzes et al. (in prep.) and Lanza et al. (in prep.). A similar configuration have been discussed in the case of Alpha Cen Bb where the stellar activity signal is larger than the planetary one \citep{2012Natur.491..207D, 2013ApJ...770..133H}.\\
 
By observing in the infrared, SPIRou is expected to be less sensitive to stellar activity since the contrast between the photosphere and the spots is smaller in the infrared than in the optical \citep{2006ApJ...644L..75M, 2008ApJ...687L.103P}. A few observations with SPIRou will therefore help to better model stellar activity and better constrain planetary masses detected with current spectrographs (e.g. HiReS, HARPS, HARPS-N, SOPHIE) or future ones (e.g. ESPRESSO, APF). Furthermore, since SPIRou is designed as a spectropolarimeter to study the magnetic fields of stars, it will be a powerful facility to understand stellar activity of stars \citep[e.g.][]{2008MNRAS.390..567M,2010MNRAS.407.2269M,2011MNRAS.418L.133M} and to constrain exoplanetary masses orbiting actives FGKM type stars.

\section{Conclusions and discussion}

SPIRou will be a pioneer high-resolution spectropolarimeter to reach a radial velocity accuracy better than 1~m.s$^{-1}$ in the infrared (YJHK bands). This will be a cornerstone for the studies of extrasolar planets around M dwarfs (especially for those in transit) in a domain where the statistics is very low: only $\sim$ 10 low-mass planets known to orbit M dwarfs, including 3 in transit. The new SPIRou instrument will have strong synergies with current and future photometric ground-based observatories (e.g. Mearth, ExTrA) and space missions (\textit{Kepler}, \textit{CHEOPS}, \textit{TESS}, \textit{PLATO 2.0}), being the most efficient spectrograph to characterize the mass of planets orbiting M dwarfs. The observations that SPIRou will perform will be extremely useful for planet formation, migration and evolution theories, as well as to provide fully-characterized planets as key targets for future atmospheric characterization with, e.g. \textit{JWST}, \textit{ECHO} from space and the E-ELT from the ground.\\

In a more general context (i.e. not limited to M dwarfs studies), SPIRou will also support the characterization of new planets orbiting actives stars that will be in the scope of spectrographs like SOPHIE, HARPS, HARPS-N, ESPRESSO, APF, etc\dots

\begin{acknowledgements}
AS acknowledges the support by the European Research Council/European Community under the FP7 through Starting Grant agreement number 239953. AS is also grateful to the administrative council of SF2A for providing him a grant to attend the 2013's annual meeting.
\end{acknowledgements}


\bibliographystyle{aa}  
\bibliography{sf2a-template} 

\end{document}